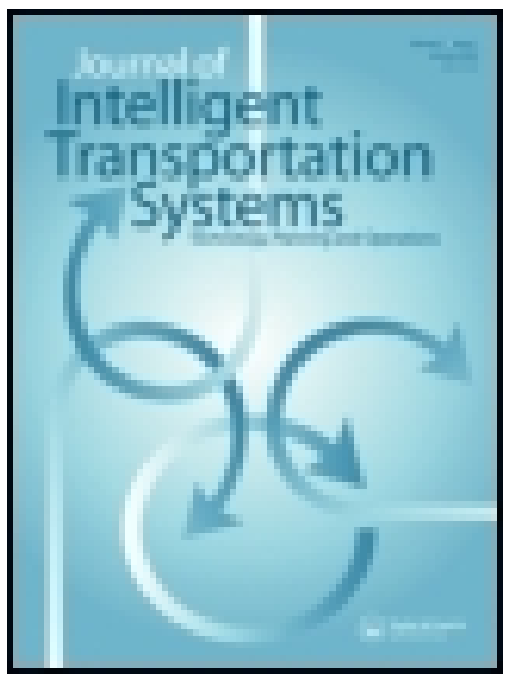

# A novel pedestrian road crossing simulator for dynamic traffic light scheduling systems

**Dayuan Tan, Mohamed Younis, Wassila Lalouani, Shuyao Fan & Guozhi Song**

# A novel pedestrian road crossing simulator for dynamic traffic light scheduling systems

Dayuan Tan[a], Mohamed Younis[a], Wassila Lalouani[b], Shuyao Fan[c], and Guozhi Song[c]

[a]Department of Computer Science and Electrical Engineering, University of Maryland, Baltimore, MD, USA; [b]Department of Computer and Information Science, Towson University, Towson, MD, USA; [c]School of Computer Science and Technology, Tiangong University, Tianjin, China

**ABSTRACT**

The major advances in intelligent transportation systems are pushing societal services toward autonomy where road management is to be more agile in order to cope with changes and continue to yield optimal performance. However, the pedestrian experience is not sufficiently considered. Particularly, signalized intersections are expected to be popular if not dominant in urban settings where pedestrian density is high. This paper presents the design of a novel environment for simulating human motion on signalized crosswalks at a fine-grained level. Such a simulation not only captures typical behavior, but also handles cases where large pedestrian groups cross from both directions. The proposed simulator is instrumental for optimized road configuration management where the pedestrians' quality of experience, for example, waiting time, is factored in. The validation results using field data show that an accuracy of 98.37% can be obtained for the estimated crossing time. Other results using synthetic data show that our simulator enables optimized traffic light scheduling that diminishes pedestrians' waiting time without sacrificing vehicular throughput.

## Introduction

Traffic signals are usually employed to ensure the safety of all users when crossing an intersection. In fact, even if autonomous vehicles become popular on the road in the future, traffic signals will continue to be the predominant means to regulate intersection crossings in order to factor in human behavior due to the presence of both conventional vehicles and pedestrians on the road (Younis et al., 2020). The increased urbanization has grown the load on road networks by both vehicles and pedestrians. Road intersections are affected the most by such high demand. The signal timing conventionally factors in a vehicle's interest in terms of reduced waiting time, and system's metrics, namely road utilization and vehicular throughput. Hence, a traffic signal controller plays a major role in managing traffic flow. The notion of intelligent transportation systems (ITS) refers to the integration of advanced electronics and information technology to enable the autonomous operation of vehicles and efficiency of roads (Dimitrakopoulos & Demestichas, 2010). Dynamic adjustment of signal timing is one means for achieving the ITS objective.

The pedestrian crossing could conflict with the objective of sustaining high vehicular throughput and low vehicle waiting time. No wonder, road management in general and traffic signals in particular, often overlook the effect of pedestrians. For example, numerous techniques have been proposed to dynamically adjust traffic signal timing and phases to cope with real-time changes in vehicle density. Yet, the size of the crowd at intersections is not factored in, causing increased inconvenience and motivating aggressive behavior by impatient pedestrians who take risks and unsafely cross the roads before vehicles stop. According to the US Center of Disease Control and Prevention, about 7700 pedestrians were killed in traffic crashes, and an estimated 182,000 pedestrians were treated in emergency rooms for nonfatal crash-related injuries in 2019 (Centers for Disease Control & Prevention, 2021). In Europe, pedestrians account for about 27% of the road fatalities (Pala et al., 2021). Published studies have pointed out that pedestrians

**CONTACT** Dayuan Tan dayuan1@umbc.edu Department of Computer Science and Electrical Engineering, University of Maryland, Baltimore County, 1000 Hilltop Circle, Baltimore, MD 21250, USA.

are 1.5 times more likely than passenger vehicle occupants to be killed in a car crash (Beck et al., 2007). Most crashes involving pedestrians happen at road crossings (Uttley & Fotios, 2017).

Such alarming statistics motivate the need for studying pedestrian behavior before and during the crossing, and what factors affect crossing decisions. Numerous studies have been conducted to simulate pedestrian behavior using mathematical and empirical models (Feliciani & Nishinari, 2016; Kudinov et al., 2018), or using virtual reality technology that mimics emulated scenarios and observes participant responses (Pala et al., 2021). However, almost all existing simulators are designed for non-signalized crosswalks (Kang et al., 2022; Wu et al., 2022). Developing a simulation model that estimates the pedestrian waiting time and the time needed for road crossing at a fine-grained level is indeed needed. Accurate assessment of pedestrian-related metrics will enable dynamic adjustment of traffic light scheduling to provide more frequent crossing phases and allot sufficient crossing time, and consequently risky behavior will be unwarranted and no pedestrian gets trapped in the middle of the road during the crossing. This paper fills that gap and makes the following contributions:

- Develop a novel Pedestrian Road Crossing Simulator (PCS), which can be used for estimating the needed crossing time at signalized intersections. PCS accurately models pedestrian motion patterns and is made available to the technical community as open source.
- Conduct a field study to validate the effectiveness of PCS. The validation results have indicated an accuracy that exceeds 98%, compared to real measurements.
- Provide a detailed case study where PCS is used to enable pedestrian-centric fine-grained scheduling of traffic signals. Multiple experiments have been conducted using both real-world and synthetic traffic demands. The results have confirmed the advantage of PCS, where the signal schedule could be adjusted to improve pedestrian waiting time while sustaining high vehicular throughput.

The article is organized as follows: "Related work" section summarized existing pedestrian crossing simulators. Our novel pedestrian road crossing simulator is presented in "Simulator for pedestrians road-crossing" section. "PCS validation and utility" section validates PCS and demonstrates how it can be incorporated to optimize traffic light scheduling (TLS) systems in a real-world setting. A use-case of PCS in TLS design is implemented and its performance has been assessed in "Using PCS in TLS design" section. "Conclusion" section concludes our article.

## Related work

Most of the existing TLS systems focus on reducing the average waiting time of vehicles and maximizing the vehicle throughput at intersections, without considering pedestrians. The main challenges addressed by these systems are: (1) how to traffic information from multiple adjacent intersections, (2) how to handle complex scenarios, for example, involving traffic in multiple directions. Aleko et al. dynamically synchronize multiple traffic lights at consecutive junctions to reduce traffic congestion based on data obtained from vehicles waiting at each intersection (Aleko & Djahel, 2020). An adaptive TLS system for a mixed vehicular scenario with human-driven vehicles, autonomous vehicles, and vehicle platoons has been developed in Tan et al. (2021). However, as pointed out above, pedestrians are often being overlooked and not factored in the optimization of traffic signal scheduling. On the other hand, quite a few studies of pedestrians' road-crossing have been conducted in recent years in order to understand behavior and provide support for safety. Some of these studies are motivated by advances in autonomous self-driving vehicles and opt to understand the implications and complications of the pedestrian–vehicle interaction in the absence of drivers. Published studies either focus on spatial models to predict pedestrian movement, or employ technologies to conduct experiments in emulated road setups.

To capture pedestrian movement, a spatial model simply tiles the crosswalk with a certain pattern in order to simplify tracking a human progression from one side of the road to the other. For example, Wang et al. have used a Cellular Automata (CA) based model to simulate how pedestrians' crossing decisions influence vehicles' travel time on road segments (Wang et al., 2018). Feliciani et al. have improved such a spatial model by combining CA with a square-shaped grid in order to enable fine-grained simulation of pedestrian motion Feliciani and Nishinari (2016). They have conducted a field experiment by monitoring a non-signalized crosswalk in the city of Milan, Italy using a surveillance camera (Gorrini et al., 2016). Their spatial model has been further used to assess the safety and estimate yearly pedestrian fatality (Feliciani et al., 2020) and to evaluate the necessity of adding a new crosswalk or installing traffic lights (Feliciani et al., 2017). Besides, a hexagonal grid has been pursued (Kudinov et al., 2018; Smirnov et al., 2020). The advantage of such a hexagon style of tiling is that pedestrian walking directions would be either straight, or $60^{\circ}$ to left or to the right, instead of turning at $90^{\circ}$ when using a square-shaped grid. However, existing spatial models do not accurately reflect practice where pedestrians move in directions with any degree during congestion.

On the other hand, some work has relied on various technologies to emulate the road environment and to study the pedestrians' behavior during the crossing. Most of the published studies have focused on non-signalized intersections. Virtual reality (VR) technology, for example, cave automatic virtual environment (CAVE) and head-mounted displays, are used to conduct experiments and study how pedestrians make crossing decisions (Mallaro et al., 2017; Pala et al., 2021). While the aforementioned experiments have been conducted off-road and rely only on standard VR equipment, Feldstein et al. have conducted field studies and built a specialized lab to better mimic practical scenarios (Feldstein & Dyszak, 2020). Similarly, Cavallo et al. have used ten movable large screens to build a CAVE-like virtual street-crossing simulator to study how older pedestrians will react to non-signalized crossing (Cavallo et al., 2019).

Some of the published studies are geared to capture vehicle–pedestrian interaction using either spatial models or special tools. Helbing et al. model both vehicles and pedestrians as moving particles to study how they influence the motion of one another (Helbing et al., 2005), where pedestrians look for a sufficient gap to cross the road within the vehicle stream at non-signalized crosswalks while vehicles opt to avoid stoppage. Both careful/aggressive drivers, and cautious/daring pedestrians are considered. Griffiths uses a mathematical model for non-signalized pedestrian crossing to derive an explicit formula for the vehicle average queue length (Griffiths, 1981) since vehicle traffic is forced to stop when encountering pedestrians. Obeid et al. utilize the DriveSafety's DS-600c Research Simulator located in the Transportation and Infrastructure Laboratory of the American University of Beirut (Obeid et al., 2017). Their study focuses on the driver-pedestrian interaction from the drivers' perspective. A system for Autonomous Pedestrian Crossing (APC) protocol of non-signalized intersections is designed in El Hamdani et al. (2020) to decrease the vehicle's delay and pedestrian's walk duration. Waizman et al. use their SAFEPED simulator to study pedestrian and driver interactions under the traffic situation of potential dangers and estimate accident risks (Waizman et al., 2015).

Little attention has been paid to modeling pedestrian's behavior while crossing on signalized crosswalks. Hashimoto et al. have studied pedestrian crossing decision and behavior at intersection crossing and how it relates to a possible collision with turning vehicles (Kamal et al., 2015). A dynamic Bayesian network is proposed to model the pedestrian decision, while factoring in many contextual parameters such as signal status, pedestrian position, motion type, etc. PCS opts to estimate the needed crossing time, rather than predict pedestrian decisions and intentions. Overall, almost all the existing pedestrian road-crossing studies are for non-signalized crosswalks and consider individual pedestrians without factoring in practical issues related to human–human interaction. Particularly, the behavior of pedestrian groups is not captured, including how individuals react when encountering other people while crossing, the needed time for groups of pedestrians to cross the road, and so on. Moreover, signalized intersections have not received much attention where the estimated time for pedestrian groups to cross a road can help the traffic lights controller in optimally scheduling phases and setting the green duration. In this way, pedestrians can also be taken into consideration when their safety and convenience are increased. Our proposed PCS for signalized crosswalk opts to fill such a technical gap.

## Simulator for pedestrians road-crossing

Our proposed Pedestrian's Road-Crossing Simulator (PCS) opts to provide a precise estimate of a pedestrian road crossing time. Various pedestrian types are considered. Furthermore, PCS supports a high degree of customization, making it suitable for different scenarios. Our simulator is also open source (Dayuan, 2022). This section introduces the pedestrian model we built to simulate the practical scenario where a group of pedestrians cross an intersection. The objectives are to (1) design a mechanism to reflect how humans move within a crowd, (2) capture the effect of mobility constraints, for example, children, elders, wheelchairs, etc., and (3) make the model customizable so that it can support further research on the pedestrian crossing. The parameters and the flow of operation of the PCS framework are articulated in Figure 1.

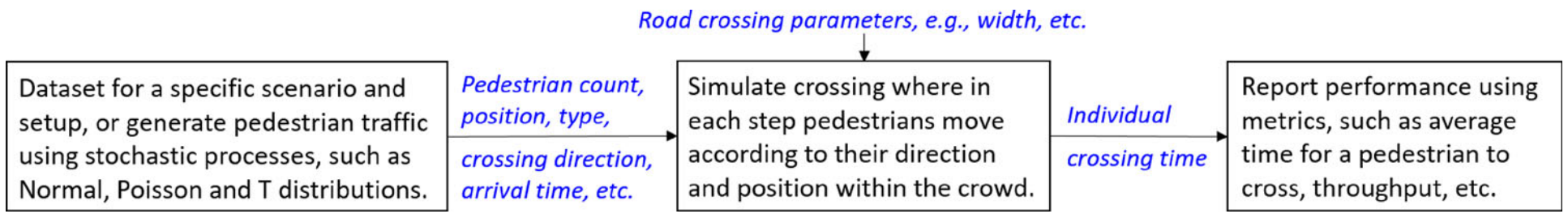

**Figure 1.** A summary of the overall PCS framework.

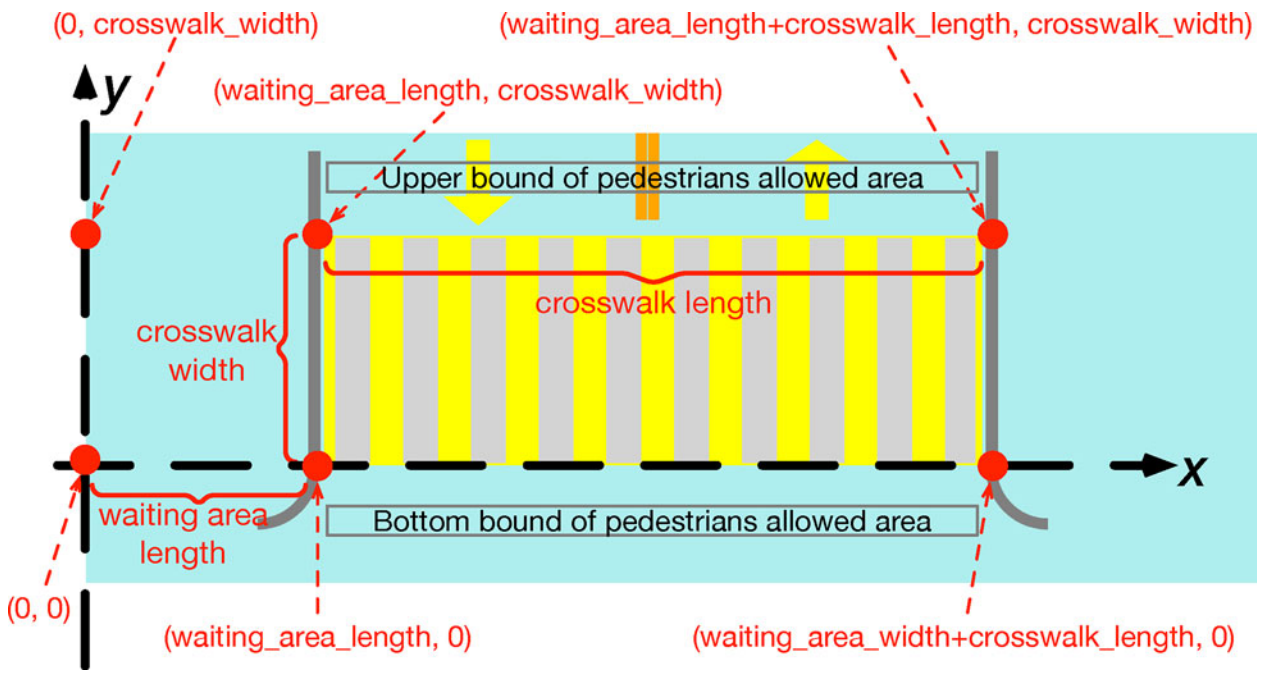

**Figure 2.** The Cartesian coordinate system.

## *Defining crosswalk area*

Figure 2 shows the layout of the crosswalk area in our simulator; the coordinates of all pedestrian positions and key points are set based on the Cartesian coordinate system shown in the figure. There are four important parameters, namely, waiting area length $\beta$, crosswalk width $W$, crosswalk length $L$, and buffer zone width $\delta$. The crosswalk is marked by the four vertices $(\beta,\ 0)$, $(\beta + L,\ 0)$, $(\beta,\ W)$, and $(\beta + L,\ W)$. There are two pedestrian waiting areas, which are on the left and right sides, respectively. There is no limitation on the number of pedestrians in the waiting area, yet the initial standing positions of pedestrians relative to the crosswalk obey specific distribution, for example, *Normal* (*Gaussian*), as we explain below.

As shown in Figure 2, there are two buffer zones above and below the crosswalk. These buffers reflect the safety area around the crosswalk; outside these buffers a pedestrian will be on the travel paths of vehicles and consequently at risk of injury. Pedestrians sometimes use these buffers, especially at crowded intersections. The coordinates of the two end-points of the upper buffer zone are $(\beta,\ W + \Delta)$ and $(\beta + L,\ W + \Delta)$, and the corresponding points for the lower buffer zone are $(\beta,\ -\Delta)$ and $(\beta + L,\ -\Delta)$. Pedestrians are only allowed to walk inside the rectangular area that combines the crosswalk and buffer zones. The buffer zones concept is derived based on our observation about what people really do in practice based on our field study. Such a concept has also been validated in our validation experiments in "PCS validation and utility" section.

## *Pedestrian parameters*

The standard waiting areas are the two areas whose vertices are $\{(0,\ 0),\ (\beta,\ 0),\ (\beta,\ W),\ (0,\ W)\}$ and $\{(\beta + L,\ 0),\ (2\beta + L,\ 0),\ (2\beta + L,\ W),\ (\beta + L,\ W)\}$, respectively. The distribution of pedestrians within the waiting area could be provided by the user as a dataset, or configured to follow any desired distribution. In the presentation in this section, it is assumed that pedestrians spread in the waiting area according to a *Normal* distribution $N(\mu, \sigma)$, which is widely used to capture human behavior in real-life applications (Martin, 2000). The mean of the such distribution corresponds to the midpoint of the crosswalk, that is, at points $(\beta,\ W/2)$ and $(\beta + L,\ W/2)$, meaning that it is highly likely that the pedestrian will be standing closer to the center points, especially when the intersection is not crowded.

If the standing positions of pedestrians are known, the PCS just uses them. Otherwise, the selected statistical distribution is used, for example, *Normal* distribution, to determine where pedestrians stand. In the latter case, we assume that each pedestrian has a probability of 95.5% to stand inside the waiting areas (dark blue shadow in Figure 3) while 4.5% to stand outside. This assumption is derived from our observation in our field study about what people would really do in this scenario. This is equivalent to having the width of the crosswalk to match $4\sigma$ for the pedestrian position $y$ coordinate distribution and having waiting area length to match $2\sigma$ for the $x$ coordinate *offset* when using $x = \beta - |offset|$ to calculate $x$ coordinates at the left waiting area and using $x = \beta + L + |offset|$ to calculate $x$ coordinates at the right waiting area. Hence, the distributions of pedestrian coordinates in the waiting area are $y$ in $N(W/2, W/4)$ and $x$'s *offset* in $N(\beta, \beta/2)$. Figure 4 shows the *Normal* distribution for 95.5% ($4\sigma$) and 99.7% ($6\sigma$). The pedestrian's position distribution is illustrated by the medium blue shadow area in Figure 3.

The simulator differentiates among the capabilities of various pedestrians. A pedestrian could be a healthy adult, an elder, a child, a disabled person using crutches, or a wheelchair. The motion speed for each pedestrian type is also assumed to be a *Normal* distribution, $N(\mu_{\text{speed}}, \sigma_{\text{speed}})$. To determine $\mu_{\text{speed}}$, and $\sigma_{\text{speed}}$, we solve the following two equations, where the maximum and minimum possible pedestrian speed are parameters and determined based on the pedestrian type.

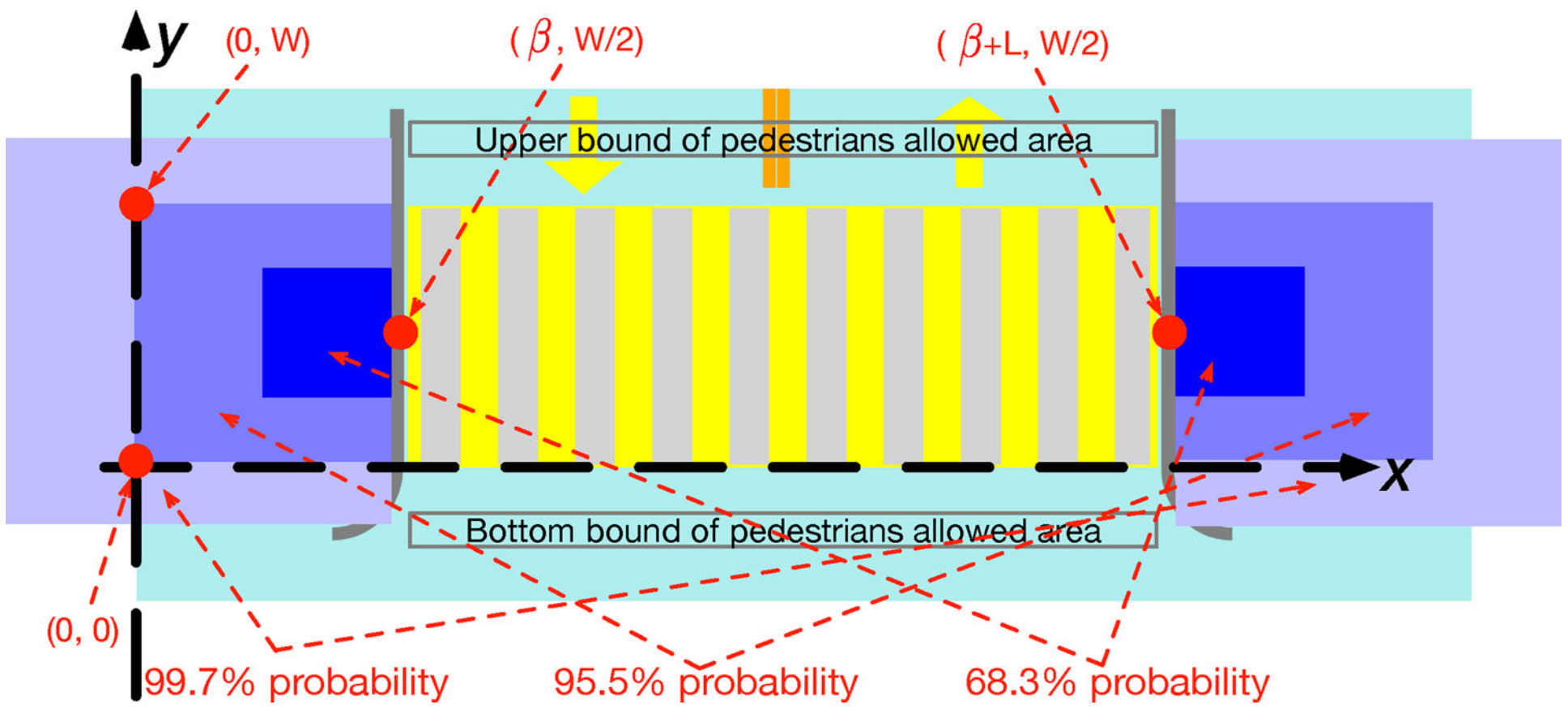

**Figure 3.** *Normal* distribution used for generating pedestrian initial standing positions.

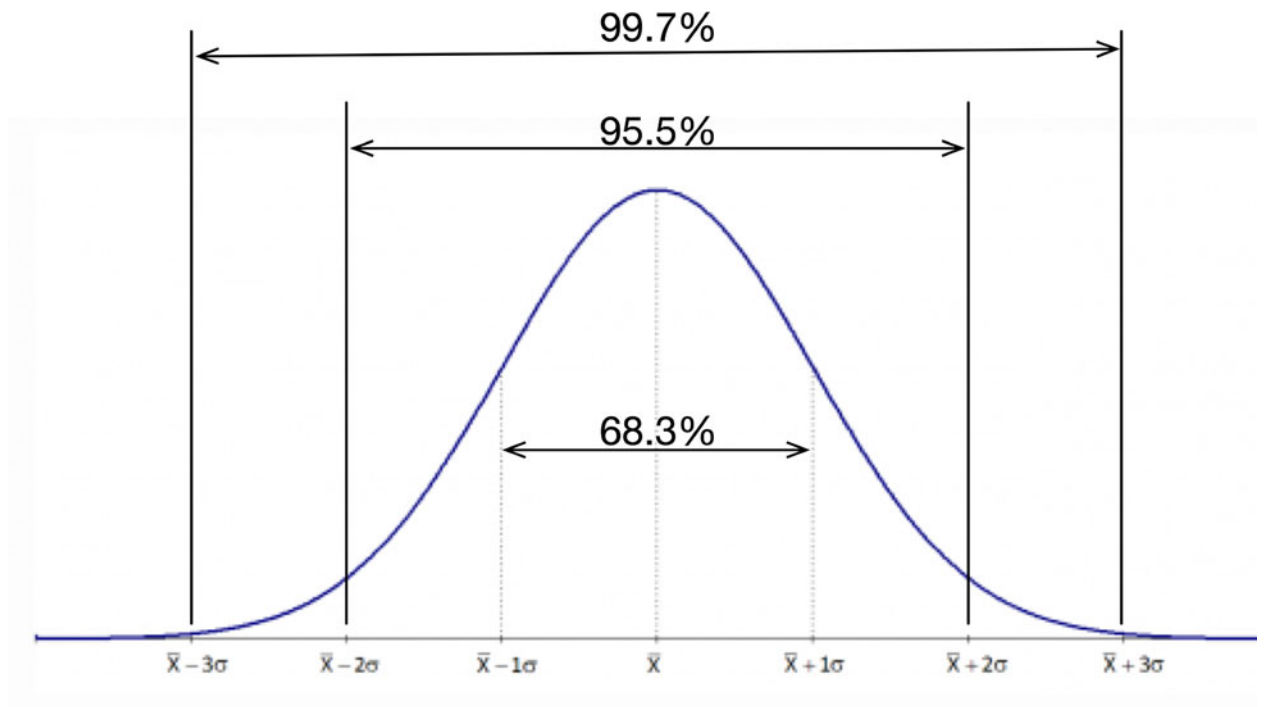

**Figure 4.** *Normal* distribution.

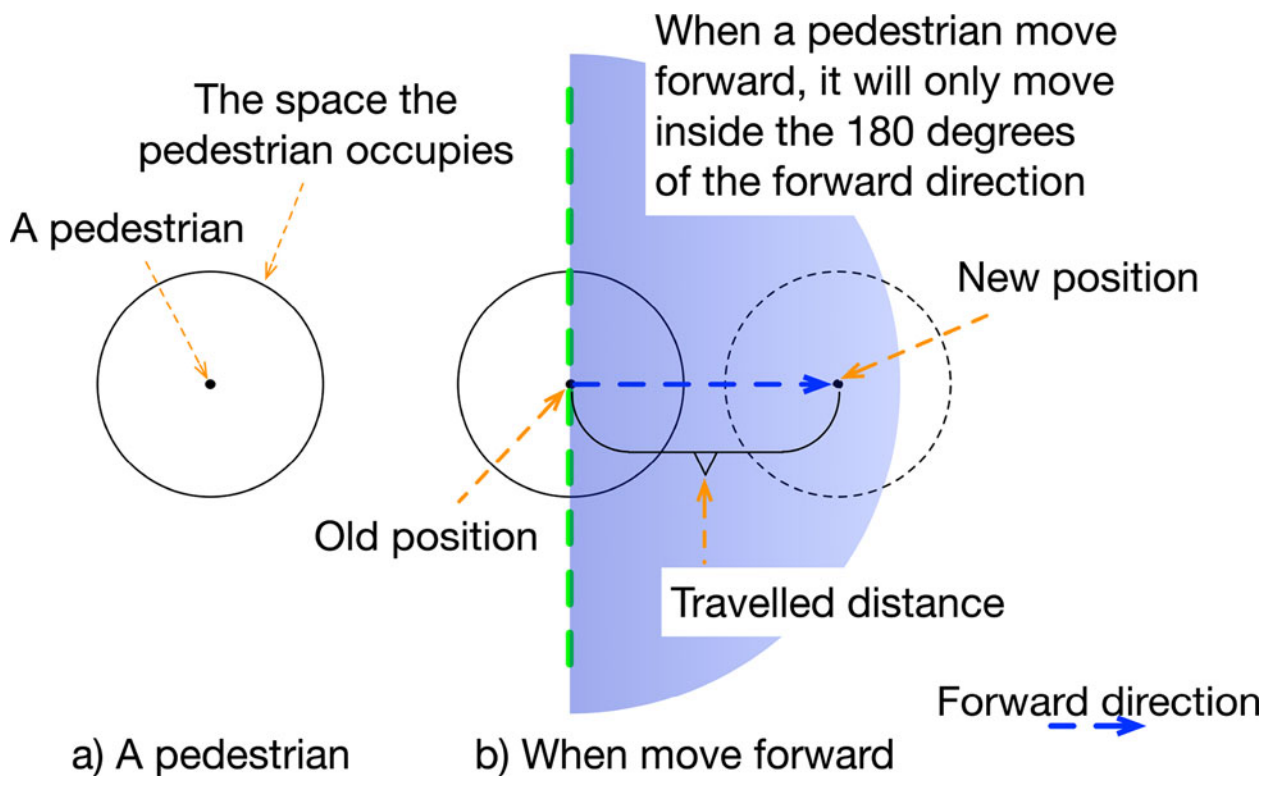

**Figure 5.** Pedestrian and its occupied space.

$$\max_{\text{speed}} = \mu_{\text{speed}} + 3\sigma_{\text{speed}} \tag{1}$$

$$\min_{\text{speed}} = \mu_{\text{speed}} - 3\sigma_{\text{speed}} \tag{2}$$

Users can also directly use $\mu_{\text{speed}}$ and $\sigma_{\text{speed}}$ as parameters if they are available. Several empirical studies have characterized the walking speed of these users, see for example "Using PCS in TLS design" section. Hence, each pedestrian has a probability of 99.7% to have a moving speed between max and min possible speeds, according to the probability density function of *Normal* distribution shown in Figure 4. We note, however, that our simulator can be easily modified to use different parameter settings if desired.

## *Pedestrian representation during motion*

In the pedestrian model of PCS, each person will be represented by a dot and the occupied space will be represented by a circle centered at such a dot with a radius that depends on the pedestrian type. Figure 5(a) shows an illustration. The circles of different pedestrians cannot overlap. When a pedestrian moves forward, the occupied circle is relocated, where the center stays on the 180 degrees ray in the motion direction, as shown in Figure 5(b). The traveled distance can be calculated using Eq. (3). The time of a simulation step can be set as any desired value; yet the default is one second. In each simulation step, every moving direction is considered independently where the farthest pedestrian from the curb moves first. All motions will be constrained by the boundary of the buffer zones as noted earlier.

$$\text{Distance} = \text{simulationstep} \times \text{pedestrianspeed} \tag{3}$$

Pedestrians always move forward as straight as possible, that is, the default direction of pedestrians is orthogonal to the curb, similar to $ped_2$, $ped_3$, and $ped_6$ in Figure 6. Only slight left or right tilts will be pursued when encountering obstacles, that is, other pedestrians. For example, $ped_4$ makes a slight left turn to avoid collision with $ped_5$. The only exception is for those pedestrians, who would be outside of the crosswalk and buffer zones, like $ped_1$ in Figure 6, where movement back to inside buffer zones is made first before resuming travel in the forward direction. When a pedestrian moves forward, the new position must be located on the half circle (illustrated as a blue dash half cycle in Figure 7) whose center is the old position and the radius is the forward moving distance. Figure 7(a) shows the case when the pedestrian's forward moving distance is larger than the radius of the occupied area circle. Figure 7(b) shows the case when it's smaller than that. The former case applies to most scenarios while the latter usually fits crutches and wheelchair users.

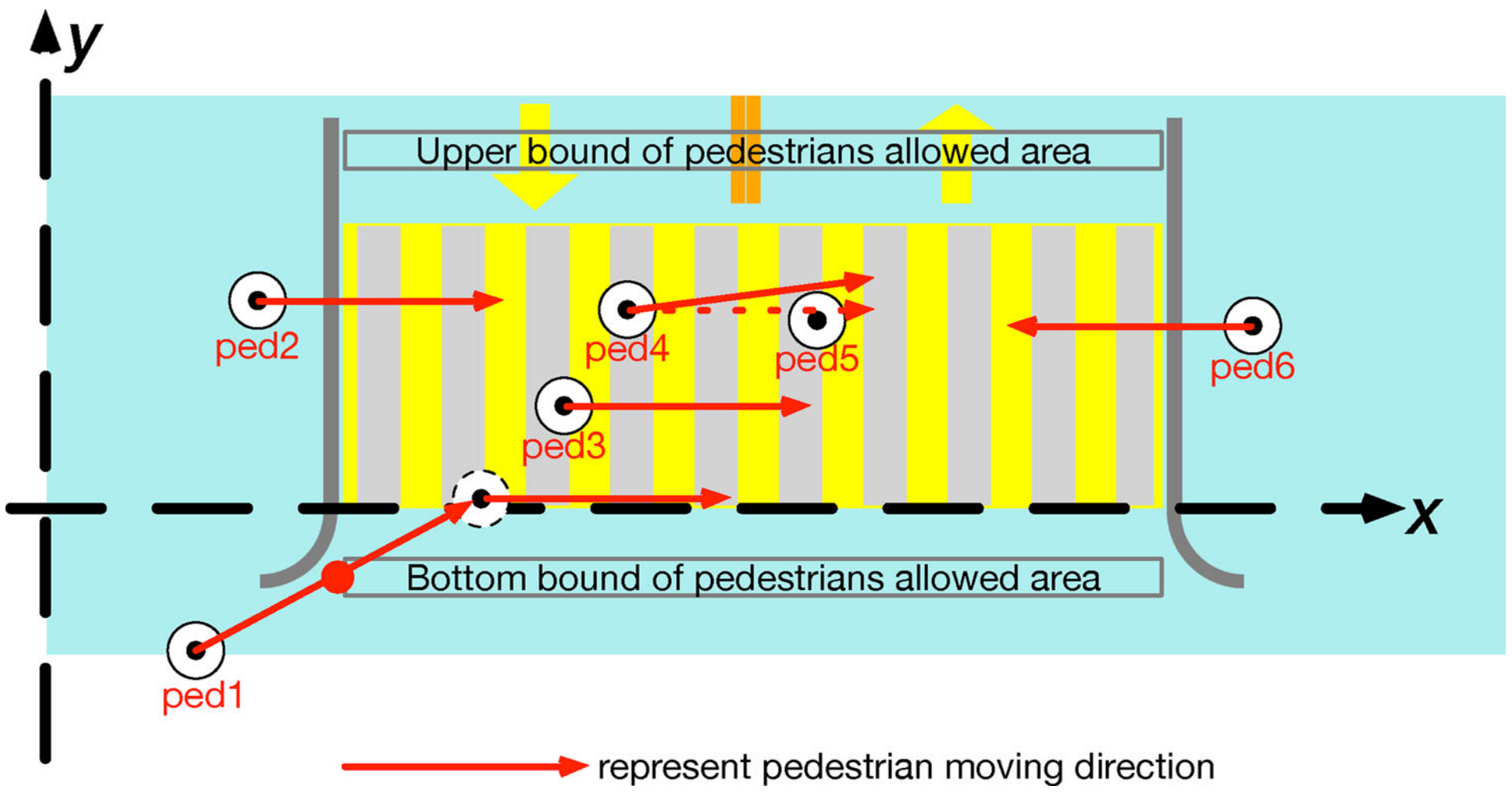

**Figure 6.** Pedestrians default directions.

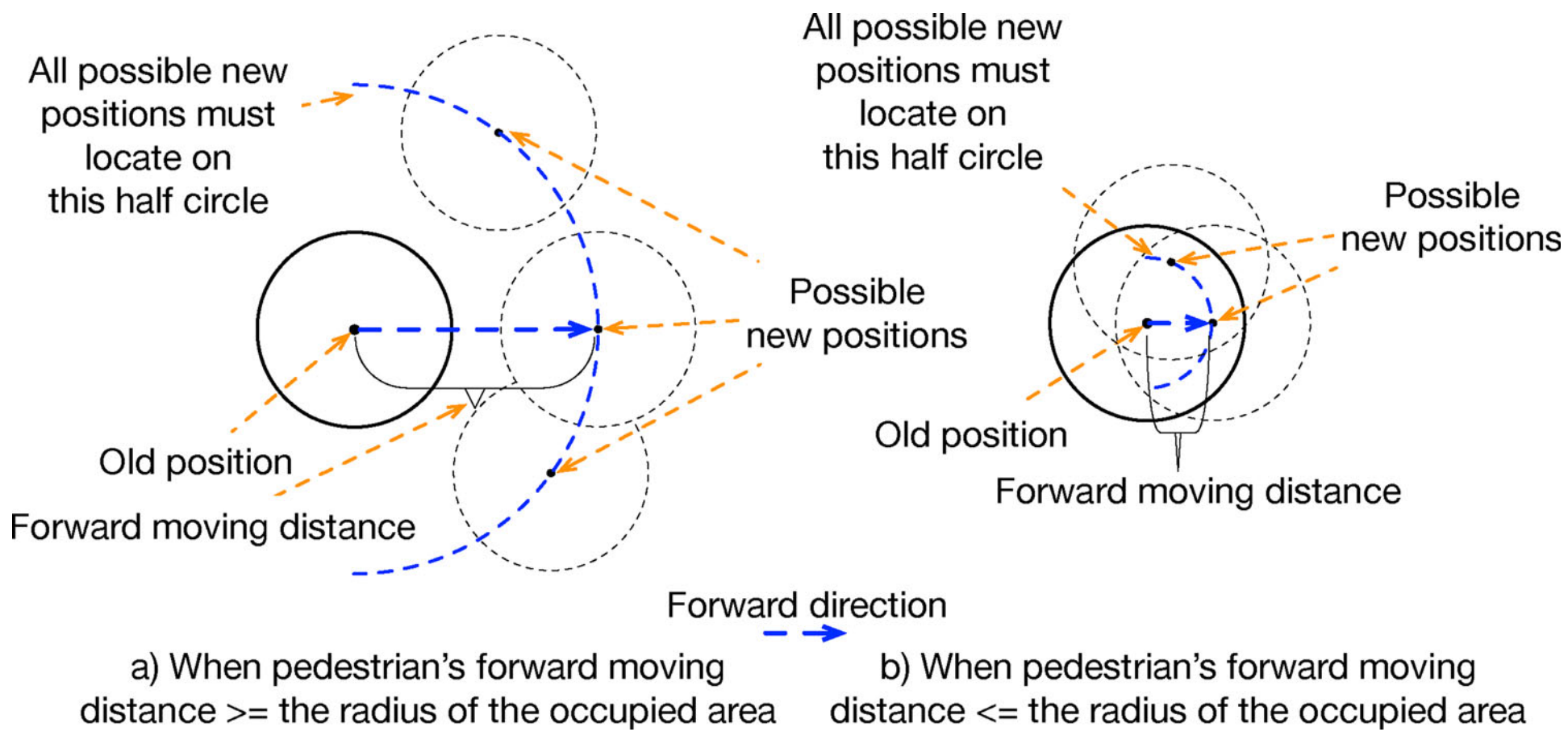

**Figure 7.** All possible new positions must be located on the semicircle whose center is the old position of the pedestrian and the radius is the forward moving distance.

Overlaps among the circles of different pedestrians are not allowed since they represent the occupied space of pedestrians in the real world. Figure 8 shows how we avoid the overlaps using an example of two pedestrians, $ped_1$ located at $(x_1, y_1)$ and $ped_2$ whose position is $(x_2, y_2)$. If $ped_1$ moves forward directly and straight to a new position $(x_1', y_1')$, $ped_1$ will encounter $ped_2$, that is, their occupancy circles will overlap. In this case $ped_1$ cannot move forward and need to tilt slightly to the left or the right. This mimics what people do in a real-life scenario. $Ped_1$ can slightly rightward to position $(x_1'', y_1'')$, or leftward to $(x_1''', y_1''')$. The circles corresponding to the two choices are both tangent to the circle of $ped_2$. The occupied circles of $ped_1$ centered at any new position located on the red arc will all overlap with $ped_2$'s circle. In conclusion, because of the existence of $ped_2$, all possible new positions of $ped_1$, represented as center dots of circles, must be located on the blue half circle excluding the red arc, a.k.a. red exclusion arc.

In a more complex scenario, multiple other pedestrians could be in the way of $ped_1$; in such a case, each of them will have a red exclusion arc with $ped_1$. Then, the union set of all those red exclusion arcs will define what cannot be $ped_1$'s new positions, as illustrated by Figure 9. Figure 9(a) shows the relative positions of four pedestrians. Figure 9(b, c) shows the red exclusion arcs of $ped_1$ caused by $ped_3$ and $ped_4$, respectively. Figure 9(d) shows the union red exclusion arc of Figures 8 and 9(b, c). Consequently, the remaining part of the blue arc is where $ped_1$'s new position can be. Obviously, the best new position should be as far away from $ped_1$'s old position in the forward direction as deemed feasible, as shown in Figure 9(d).

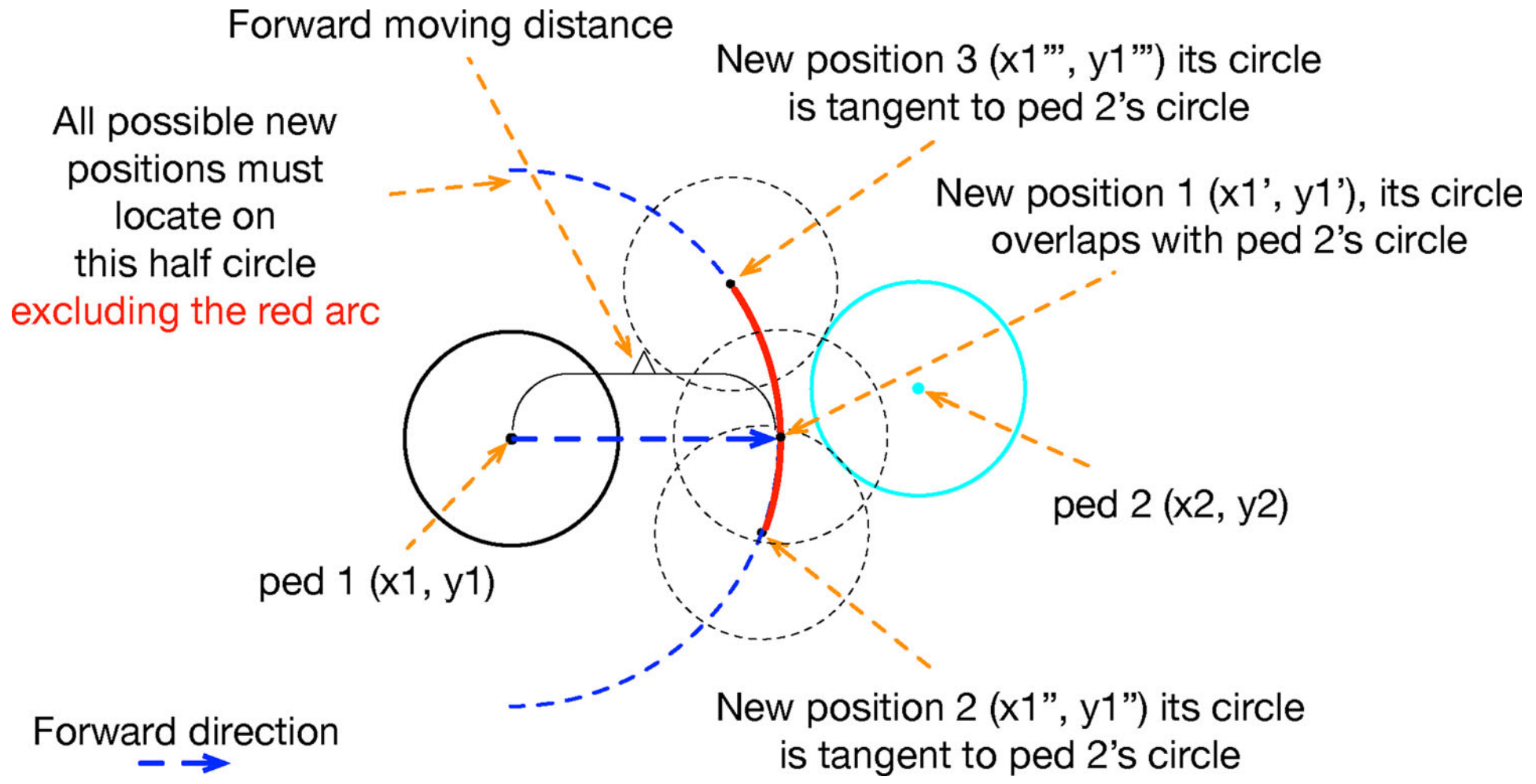

**Figure 8.** Illustrating the case when there is another pedestrian ("ped 2") in the way of "ped 1".

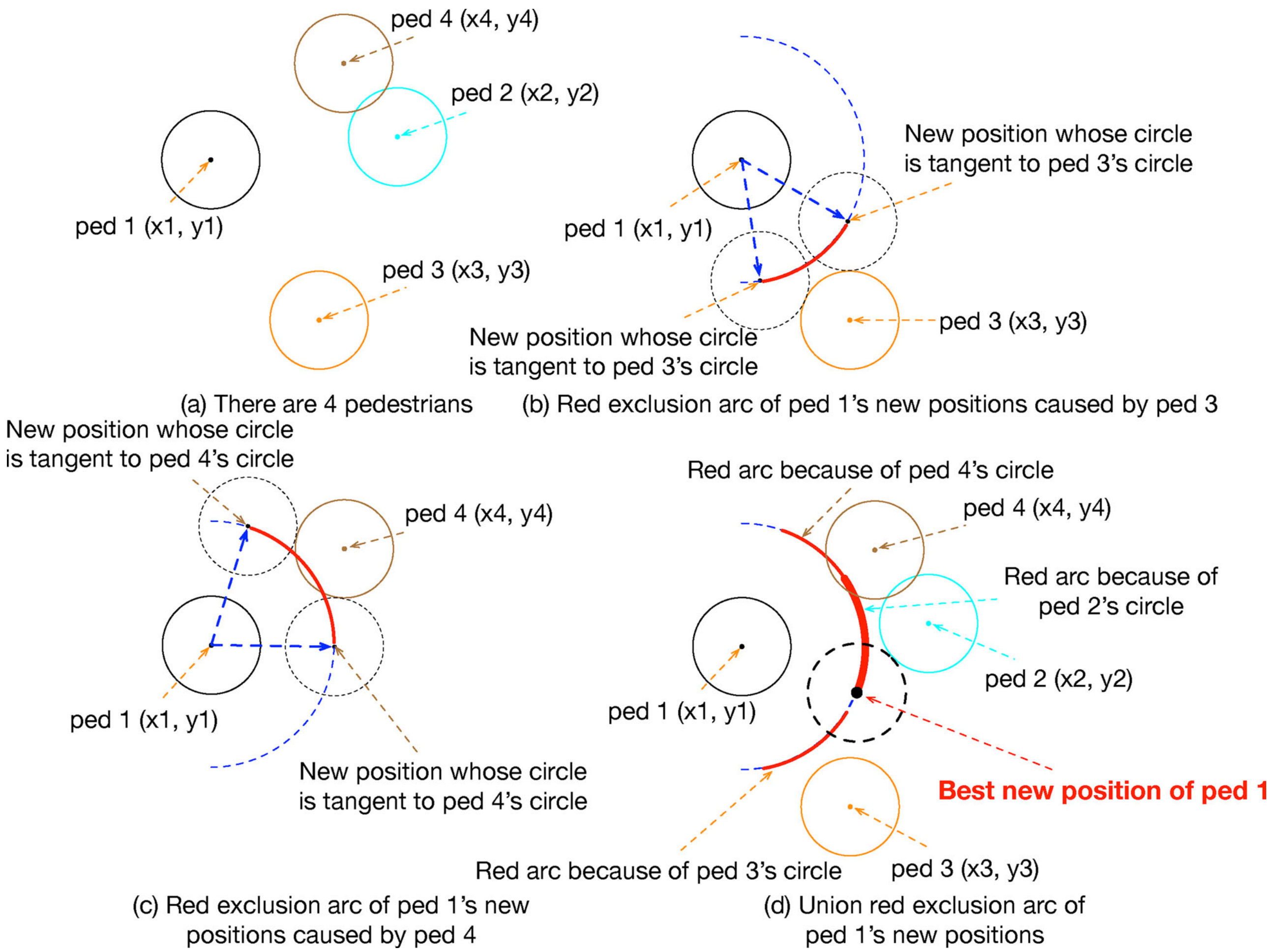

**Figure 9.** Union "red arc"/exclusion arc of $ped_1$ where $ped_1$'s new positions cannot be located when there are three other pedestrians in front of her or him.

To simulate the pedestrian's walk more accurately, the radius of each pedestrian occupied space is dynamically adjusted. Just like what people really do, when they walk faster the space they occupy is larger than when they walk slower. The maximum value of the circle radius corresponds to the pedestrian's maximum speed, while the minimum value is for when this pedestrian stands. Both values can be customized. Pedestrians will firstly try moving forward using maximum speed for a distance that can be calculated using Eq. (3). If a new position cannot be found using the method explained above, that is, the red exclusion arc totally covers the whole blue arc, then the motion speed is reduced by a certain value ($i$) using Eq. (4), where $i$ is set as a parameter supporting customization and $1 \leq i \leq 100$. The pedestrian will give up moving and stay still if a new position can be found even using minimum speed. This situation rarely happens. In our field study, discussed in "Validation using field study data" section, no-move situations have occurred less than 2% of the time. It is worth mentioning that our field study considered the most-crowded scenario. This pedestrian will try again in the next simulation step, hoping that other pedestrians move and clear the way.

Whenever a pedestrian changes the motion speed, the radius of the corresponding occupied space will be adjusted based on Eq. (5), where the $\max_{radius}$ and $\min_{radius}$ are the radius of the occupied space when this pedestrian moves with max and min moving speeds, respectively. If a pedestrian cannot move during three consecutive simulation steps, it means that the crosswalk is so crowded that it becomes blocked; then these pedestrians that got stuck in the area will all try to tilt right and move in such a direction, just like what people would do in real life.

$$\text{Speed} = \max_{speed} - \{\max_{speed} - \min_{speed}\} * i\% \quad (4)$$

$$\text{Radius} = \max_{radius} - \{\max_{radius} - \min_{radius}\} * i\% \quad (5)$$

## Best new position calculation

As noted earlier, our simulator considers pedestrians in each direction according to their progress on the road crossing. Assuming the current pedestrian who wants to move forward in the crosswalk area is $ped_1$ is located at ($x_1$, $y_1$) and has space with radius $r_1$. There are $n-1$ other pedestrians in the way of $ped_1$, whose positions are $(x_2, y_2), (x_3, y_3), \ldots (x_n, y_n)$ and occupied spaces with radii $r_2, r_3, \ldots r_n$, respectively. As shown in Figure 8, when there is another pedestrian on $ped_1$'s way, $ped_1$'s new position must be located on the half blue circle and not intersect with the red exclusion arc.

Assuming the coordinates of two endpoints of the red arc are $(x_1'', y_1'')$ and $(x_1'', y_1'')$, these two points must be the intersections of two circles, shown in Figure 10(b). The center of the first circle is centered at ($x_1$, $y_1$) and has a radius of $ped_1$'s forward moving distance, which can be calculated via Eq. (3). The center of the second circle is $ped_2$'s position and the radius is $r_1 + r_2$. Then the coordinates $x_1'', y_1'\prime, x_1'', y_1'\prime\prime$ can be calculated by solving the set of equations consisting of (6) and (7). Then, the new position $(x_{1,new}, y_{1,new})$ of $ped_1$ can be presented via the following equation, where $d_1$ is the forward distance of $ped_1$ as shown in Figure 10(a):

$$(x - x1)^2 + (y - y1)^2 = d_1^2 \quad (6)$$

$$(x - x_2)^2 + (y - y_2)^2 = (r_1 + r_2)^2 \quad (7)$$

$$\begin{cases} (x_{1,\text{new}} - x_1)^2 + (y_{1,\text{new}} - y_1)^2 = d_1^2 \\ x_{1,\text{new}} \geq x_1 \\ x_{1,\text{new}} \leq x_1'', x_{1,\text{new}} \leq x_1'' \\ y_{1,\text{new}} \geq y_1'', y_{1,\text{new}} \leq y_1'' \end{cases} \quad (8)$$

Similarly, we can get all intersections of the circle representing $ped_1$'s occupied area and the circles representing all other neighboring pedestrians. After considering all those intersections, we can achieve the overall value range of the new position $(x1_{\text{new}}, y_{1,\text{new}})$ of $ped_1$. Lastly, we pick the position $(x_{1,\text{new_final}}, y_{1,\text{new_final}})$, which has the largest $x$ value in the range, since the final picked new position for $ped_1$, which definitely is also the best choice for $ped_1$ to move forward as far as possible. The above discussion applies to pedestrians moving from the left side of the road to the right. The final picked new position of those pedestrians who move in the opposite direction can be solved in a similar way.

## Simulator interface

A Graphical User Interface (GUI) and a back-end Application Program Interface (API) are also provided. The GUI begins with a parameter settings configuration page, shown in Figure 11, which allows customization for various use cases. By defaults, pedestrian initial positions in the waiting areas are generated using a *Normal* distribution as shown in Figure 12; yet other distributions can be supported as mentioned earlier. A pedestrian occupies a smaller space when standing than when walking. Figure 13 shows one intermediate step during the simulation. While pedestrians get closer to a crowded area, if they cannot find new positions using the current occupied-space, they would slow down and meanwhile occupy less space, which makes it easier to find the next positions. The API is also provided for users who want to integrate our simulator into their projects. An example use-case is provided in Dayuan (2022).

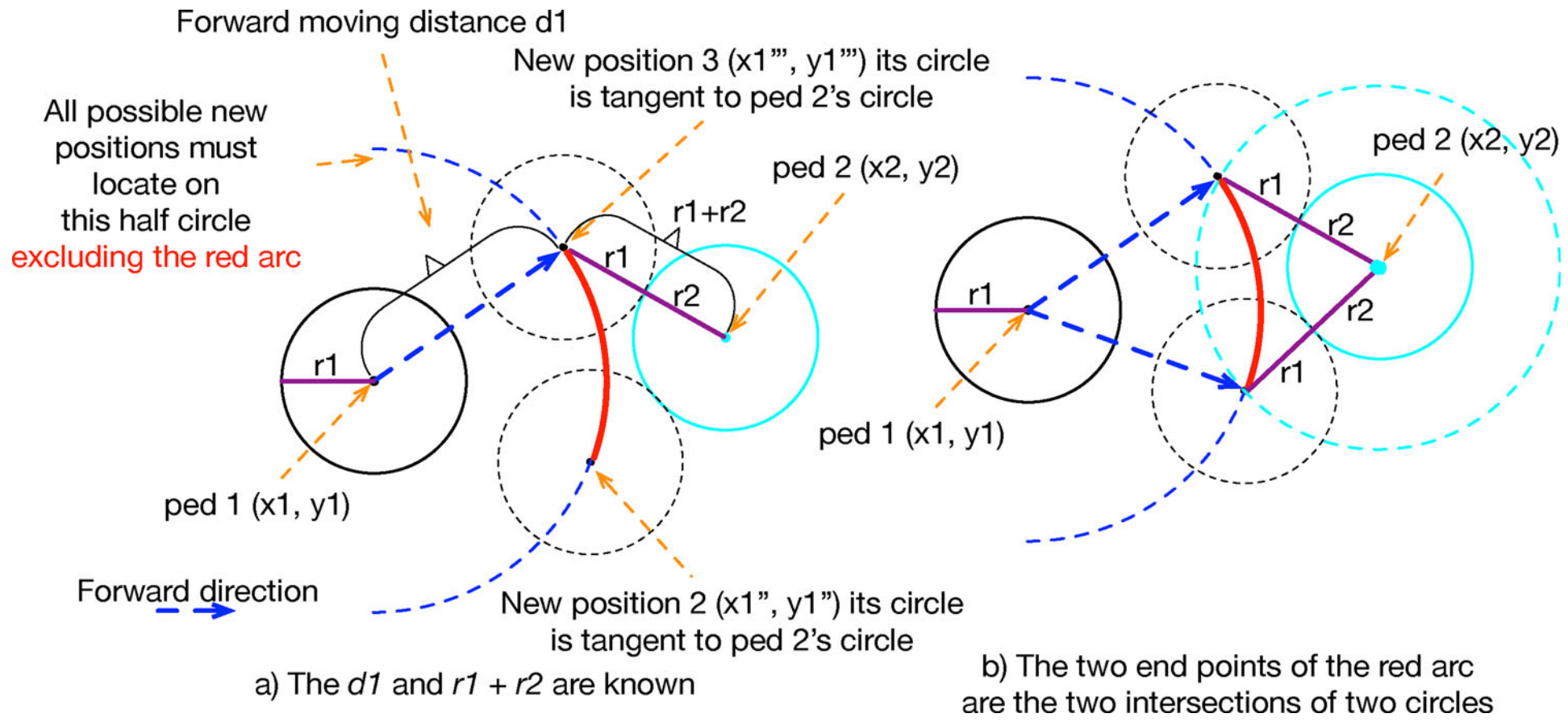

**Figure 10.** The two endpoints of the red arc are the two intersections of two circles.

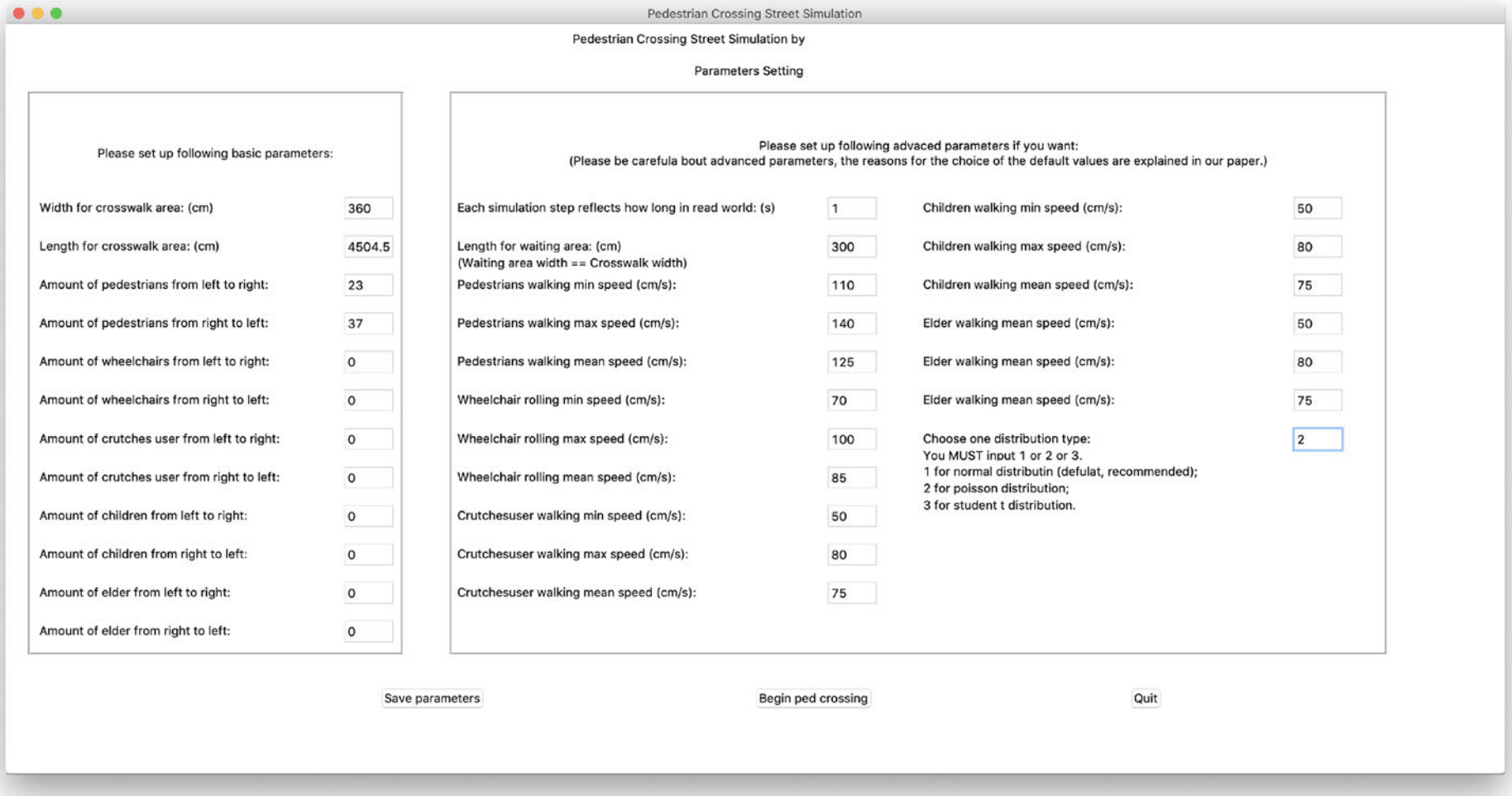

**Figure 11.** PCS can be customized by entering the parameter settings using the configuration page.

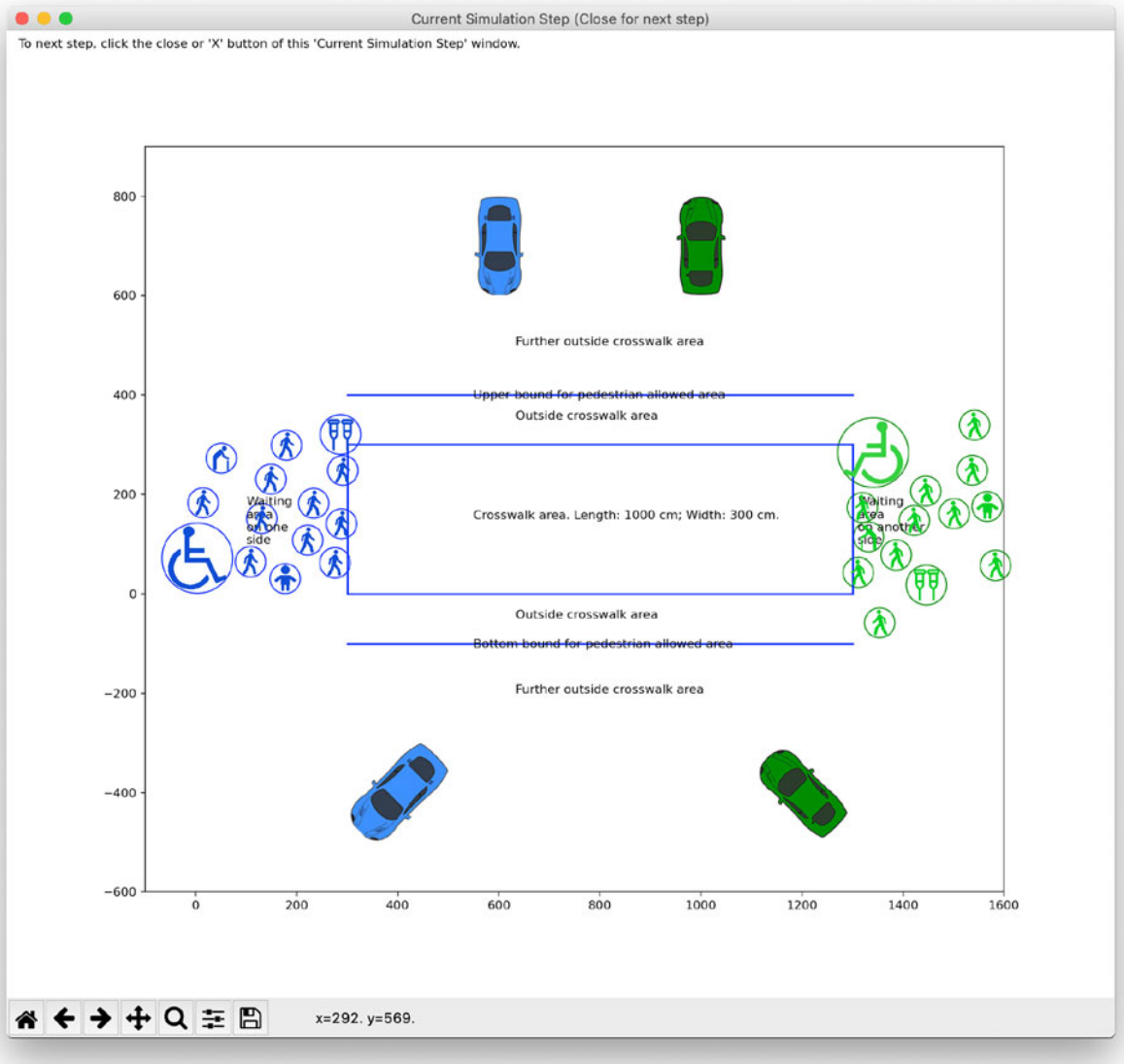

**Figure 12.** Generated initial positions for pedestrians using *Normal* distribution.

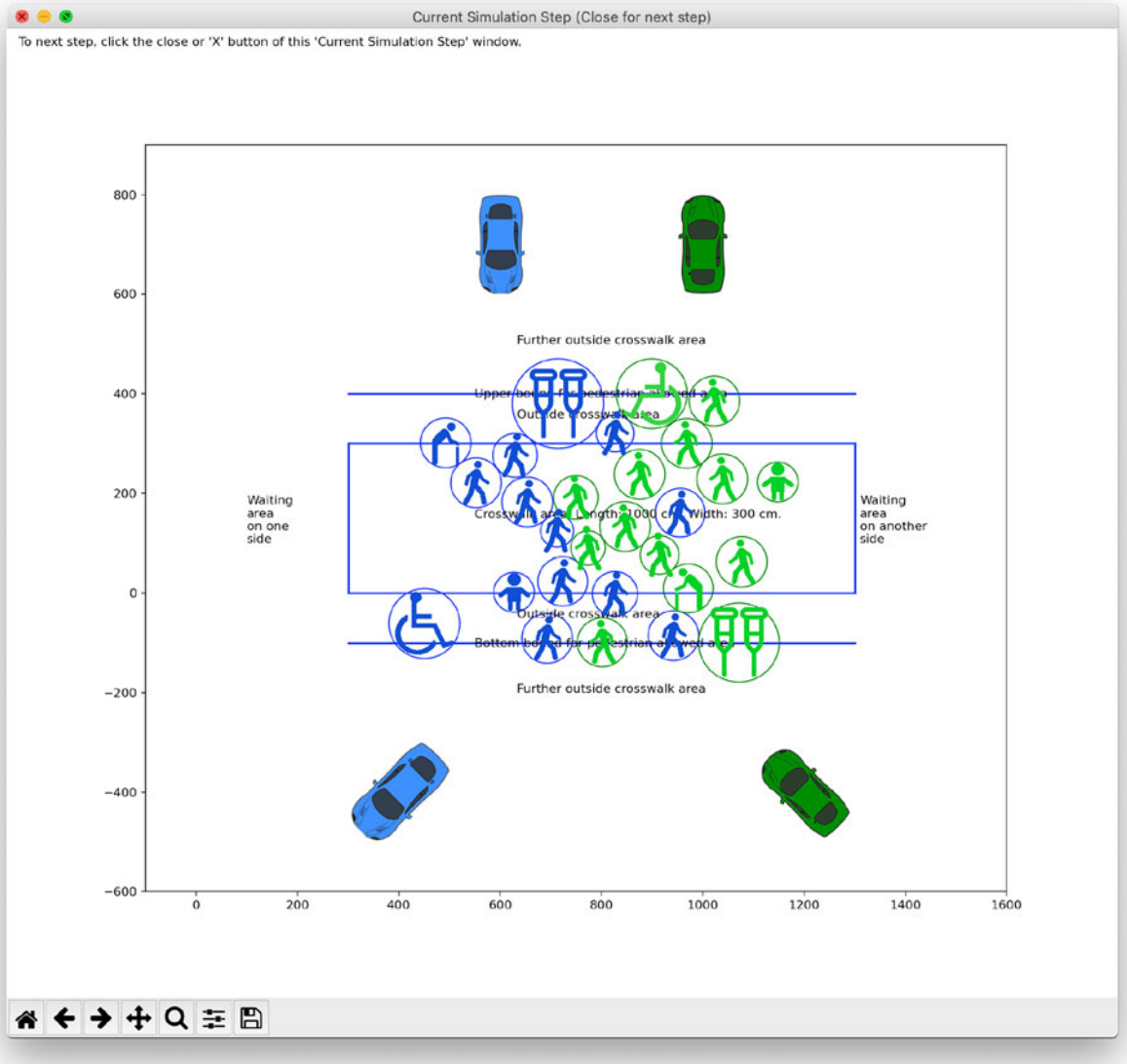

**Figure 13.** One intermediate simulation step.

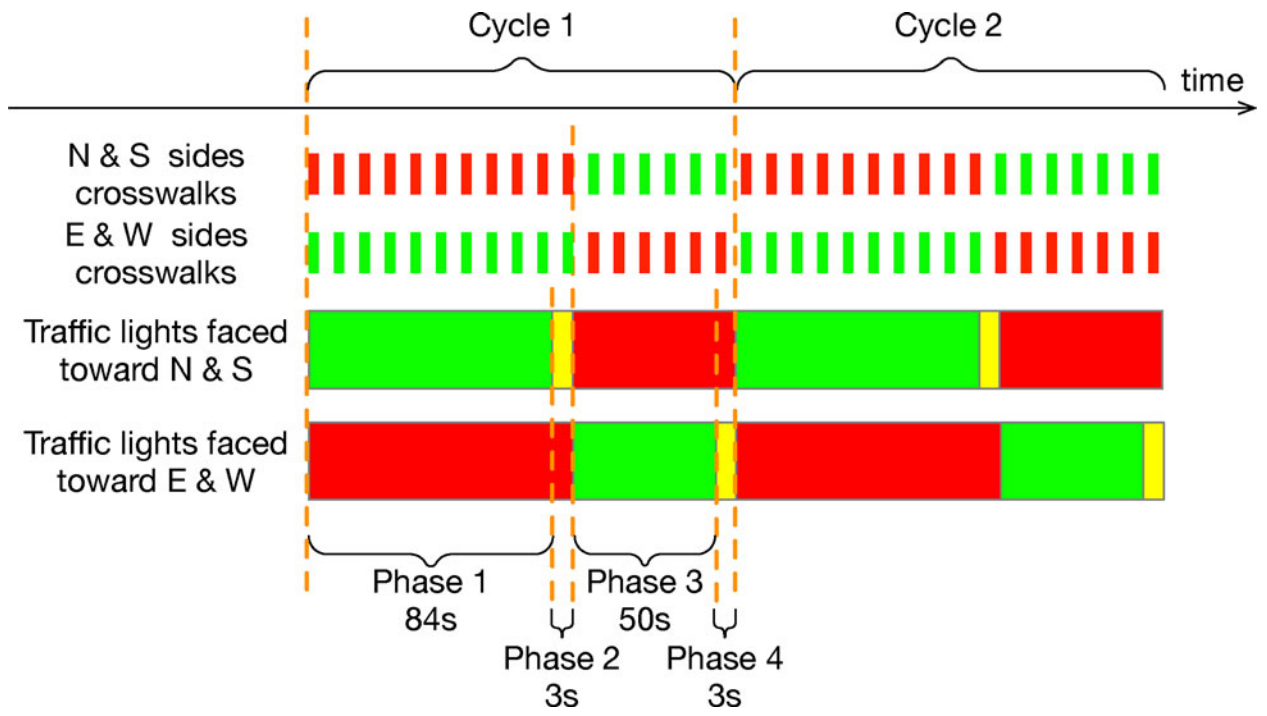

**Figure 14.** Cycle time and phases for the static signal schedule of the intersection at $Location_1$.

## PCS validation and utility

In this section, we validate the accuracy of our proposed PCS using data collected from the field. In addition, we discuss how PCS can support various traffic management studies that involve pedestrians. In particular, we present an example use-case for studying the green duration for pedestrian crossing in a dynamic TLS system. The source code for the use-case implementation is accessible at Dayuan (2022).

### *Validation using field study data*

To gauge PCS's ability to mimic practical scenarios, we have conducted a field study. The goal is to collect load and measurement statistics from the field and compare the results to what PCS generates for the same load. Multiple videos have been captured to document real scenarios via mobile phones. Vehicles, pedestrians had been counted and their directions, used time to pass had been manually recorded. All collected data is made accessible (Dayuan, 2022) so that other researchers can leverage it in conducting studies related to TLS systems, pedestrian behavior, intelligent transportation systems, etc.

#### *Monitored intersections and crosswalks*

The first field study was geared for collecting data through monitoring the intersection of Ming Guang Road and Feng Cheng No. 9 Road in Xi'an City, Shaan Xi Province, China. Such an intersection, which we refer to as $Location_1$, reflects a popular setting in a busy downtown area with four lanes on all edges. The left most lane is for turning left only. All other three lanes can be used for going straight. Only the rightmost lane allows turning right. In addition, the speed limit of incoming and outgoing edges is 60 km/h. There is one crosswalk at each side of this intersection. The length of the two crosswalks at the North and South sides of the intersection is 47.69 m; the width of both is 6.4 m. Meanwhile, the length and width of the two crosswalks on the West and East sides are 43.62 and 3.6 m, respectively.

The traffic light schedule for this intersection is fixed and is shown in Figure 14. There are four phases in each cycle. The duration of each phase is 84, 3, 50, and 3 s, respectively. Pedestrians are allowed to cross when the corresponding traffic lights for vehicles are red. For example, in Figure 14 Phase 1, pedestrians on the West and East side are allowed to cross during this phase while the West-faced and East-faced traffic lights are red. In this field study, we observed the vehicular and pedestrian traffic at $Location_1$. We recorded how many vehicles and pedestrians passed the intersection, their directions and the needed time for pedestrian road-crossing. We later used these statistics to reproduce the traffic demand and study the utility of PCS in optimizing traffic flow. We repeated the same procedure in the second field study, which involved observing the crosswalk at the north side of the intersection between Wei Yang Road and Feng Cheng No. 7 Road. Such crossroad, which we refer to as $Location_2$, has a width of 3.6 m and a length of 45.045 m. We counted how many pedestrians crossed the Wei Yang Road via this crosswalk and how much time they used.

To validate PCS, the pedestrian-road-crossing related statistics of the west side crosswalk at $Location_1$ and the north side crosswalk at $Location_2$ are used. The measurements are shown in Table 1. The first and second records were collected at around 2 pm while record 3 was at around 7 pm on Thursday, 26 August 2021, at the first crosswalk. These three observation times are carefully selected to reflect the scenario where the pedestrian traffic is heavy and the crosswalk is crowded. The fourth and fifth records were collected around 7 pm on the same day for the second crosswalk. These two data sets reflect the scenario where the pedestrian traffic is light and the crosswalk is kind of empty. In addition, thirty pedestrians were arbitrarily selected and their walking speeds were recorded. The walking speeds were found to follow a *Normal* distribution $N(1.2676, 0.09167)$.

#### *PCS validation experiments and results*

We used the collected data from our field study to validate the effectiveness of PCS. For each set of parameters, we ran our PCS ten times and calculated the average as the final results. The estimated time needed by pedestrians for road crossing has been compared with the actual time they used in practice. The results are shown in Table 1. The first, second and third records reflect the scenarios with heavy pedestrian traffic and crowded crosswalk area, while records 4 and 5 represent the scenarios with light pedestrian traffic and relatively empty crosswalk. High accuracy is observed for all five records. In addition, three distributions, namely, *Normal*, *Poisson*, and *T* distributions, have been considered for positioning pedestrians in the waiting area in our experiments, as shown in Table 1. Figure 15 presents the results for the *Normal* distribution for better visibility and comparison. These validation results closely match the observed values which demonstrate that PCS yields accurate estimates.

**Table 1.** Comparing the collected Data about pedestrians road crossing data with estimates provided by PCS.

| Records | | At $Location_1$ | | | At $Location_2$ | |
|---|---|---|---|---|---|---|
| | | Record 1 | Record 2 | Record 3 | Record 4 | Record 5 |
| # crossed pedest. | | E2W: 29 W2E: 21 | E2W: 25 W2E: 19 | E2W: 23 W2E: 37 | S2N: 3 N2S: 4 | S2N: 3 N2S: 3 |
| Actual Time (s) | | 57 | 53 | 60 | 39 | 40 |
| Norm. Dist. | Esti. Time (s) | 57.2 | 51.4 | 60.2 | 40.5 | 40.7 |
| | Accu. | 99.65% | 96.98% | 99.67% | 96.15% | 98.25% |
| Pois. Dist. | Esti. Time (s) | 56 | 55 | 61 | 40 | 38 |
| | Accu. | 98.25% | 96.23% | 98.33% | 100% | 95% |
| T Dist. | Esti. Time (s) | 56 | 54 | 66 | 38 | 39 |
| | Accu. | 98.25% | 98.11% | 90% | 95% | 97.5% |

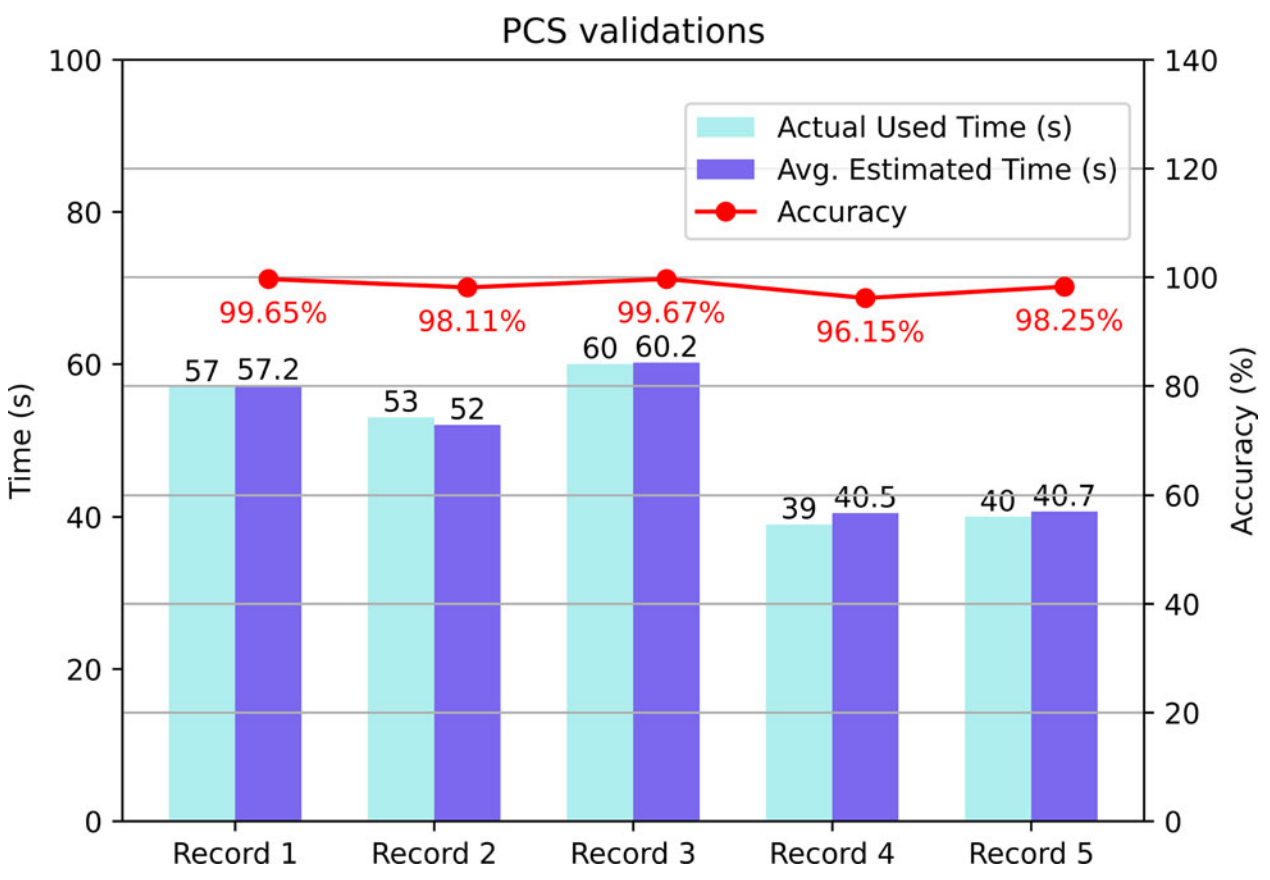

**Figure 15.** Validation results for comparing PCS (*Normal* Distribution) to actual field measurements.

## *Supporting dynamic TLS*

TLS has a well-known effect on traffic flow and trip latency. While many of the existing TLS systems are static in nature, the notion of dynamic traffic management has gained broad acceptance as an effective means for coping with vehicular density and traffic incidents in real time. In essence, static TLS systems use fixed settings for green time and phases (sequence of green signals) at intersections regardless of the traffic demand. Such a model proved to be inefficient and is being replaced by dynamic TLS that adjusts the green duration and the order of phases according to the flow in the various directions. To assess vehicular density on each road segment, a dynamic TLS leverages on-road sensors, GPS tracking, and other technologies. Such data is then fed to optimization models to find the settings that maximize throughput and minimize waiting time at the intersection (Younis et al., 2020). In fact, the scope of dynamic TLS is not confined to a single intersection, and coordinated signal timing is being pursued to improve these metrics across multiple collocated intersections (Lee et al., 2019; Tan et al., 2021).

Factoring pedestrians in the traffic management optimization has been lacking. Most existing work has focused on the vehicular traffic aspect of the performance without factoring in or even caring about the implications on pedestrians. For example, increasing the green time for one direction at an intersection could force pedestrians to wait for a long time; such excessive road crossing latency affects pedestrians' behavior, causing irrational moves and crossing violations by some impatient individuals and risking collisions or unsafe vehicle stoppage. In addition, the red time could be insufficient leaving some pedestrians in the middle of the road while allowing vehicles to pass. On the other hand, a very cautious approach would not be helpful to vehicular traffic. Basically, providing generous crossing time for pedestrians based on a prefixed assumption about the number of involved individuals would keep vehicles waiting for no one to cross. One of the main reasons for excluding pedestrians from dynamic TLS decisions is the lack of an accurate assessment framework that can gauge the implication of certain decisions on pedestrians. PCS fulfills such a need and enables the design of pedestrian-centric dynamic TLS systems. Figure 16 shows how PCS can be integrated into the dynamic management of TLS. We note that the optimization function will differ when pedestrians are factored in determining the green times and phases, as we noted above. In the next section we show how PCS can be beneficial, where we compare the performance of static and dynamic TLS systems with respect to both vehicles and pedestrians.

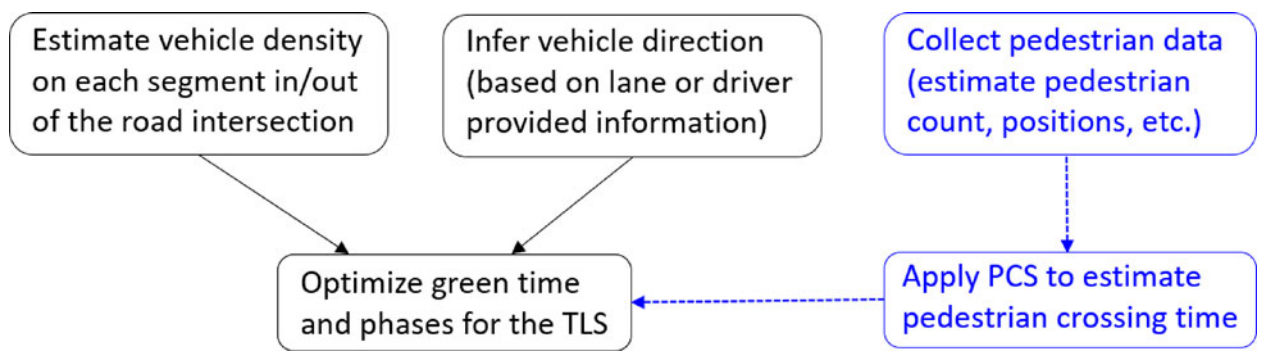

**Figure 16.** A block diagram showing how PCS can be leveraged to factor in pedestrians in dynamic TLS setting.

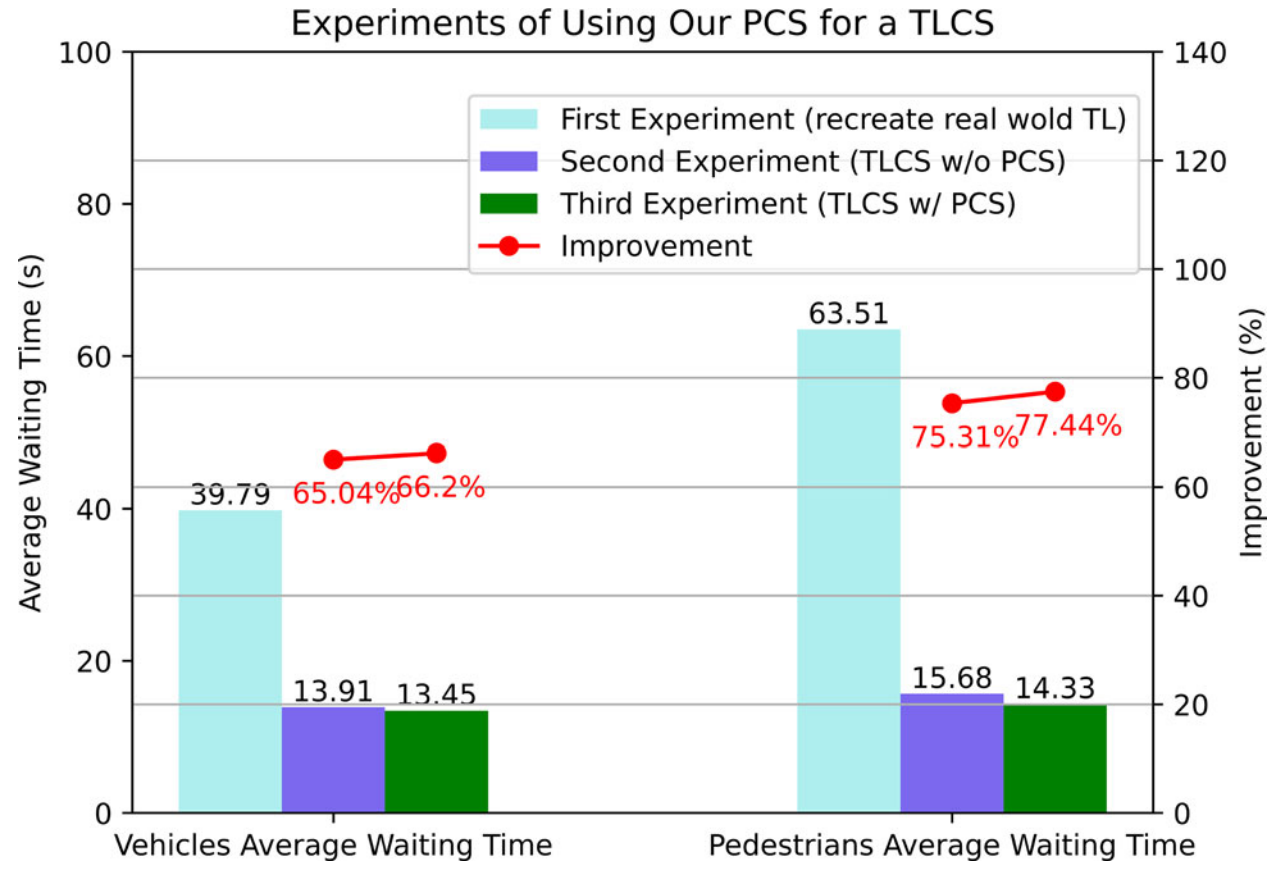

**Figure 17.** Experiments of using our PCS for a TLS system and the improvement on Average Waiting Time (AWT).

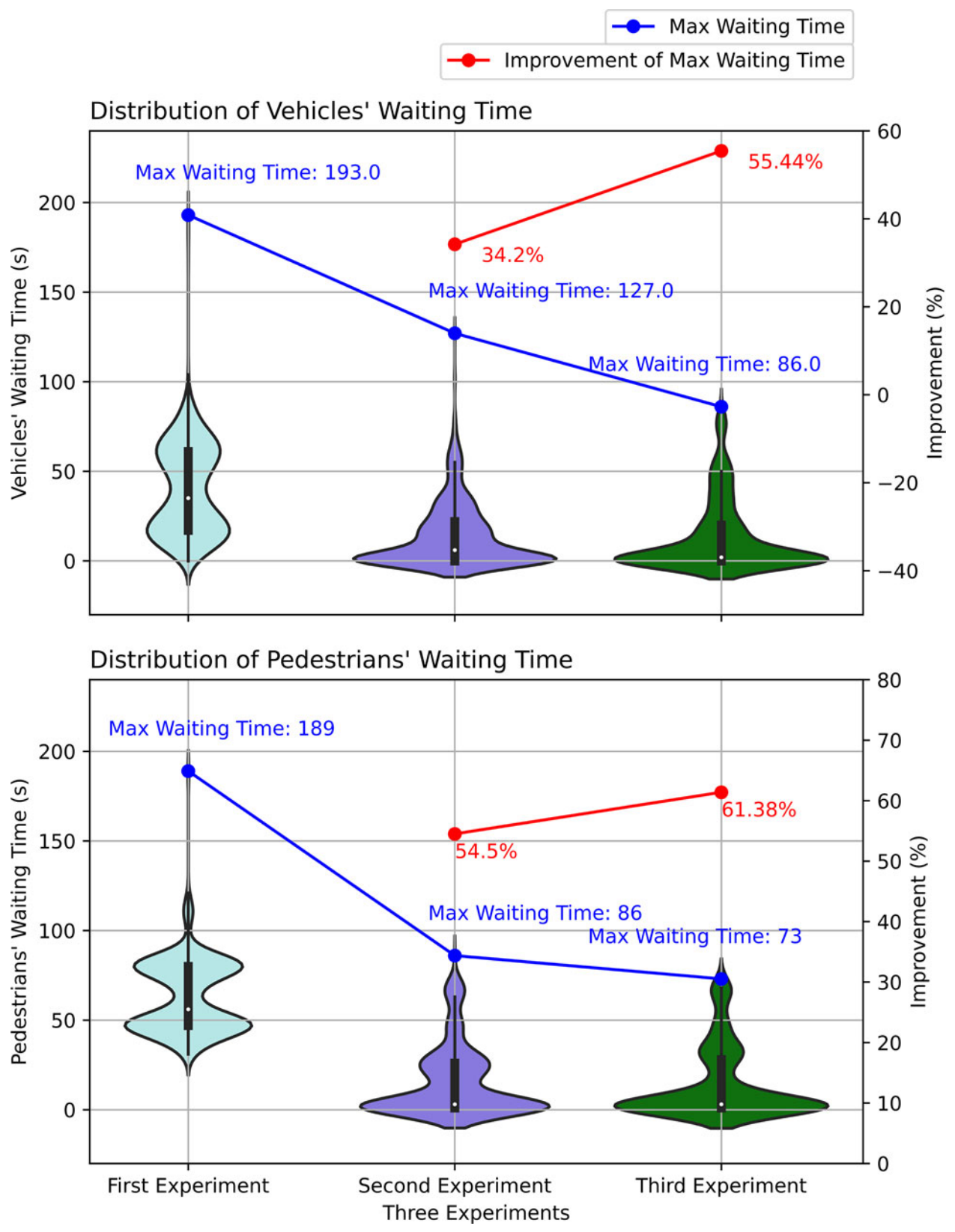

**Figure 18.** Distributions of the vehicle and pedestrian waiting time. The width of the bar reflects the frequency of occurrence (meaning how many experiences such waiting time) while the height reflects the maximum waiting time experienced by any among the considered population.

## Using PCS in TLS design

In order to demonstrate the utility of PCS, we have conducted experiments to compare various TLS design options. The experiments are based on a four-way intersection with four lanes in each direction, as illustrated in "PCS validation and utility" section. The tools we used to conduct our experiments are SUMO v1.9.2 (Lopez et al., 2018), which is widely used to simulate traffic scenarios, and Python 3.7.2, which is used to implement our algorithms and to control vehicles, pedestrians, traffic lights, and other objects in SUMO simulation.

### *Considered TLS configurations*

To compare the performance difference between using and not using PCS, three experiments have been conducted. The first experiment assumes a static schedule where the phases and their duration are fixed. The used TLS matches what is shown in Figure 14. In the second experiment, the fixed TLS is changed to be dynamic and adaptive to traffic demand. Basically, the green duration is dynamically adjusted based on how many vehicles are going to pass the intersection. The green time is set to allow all vehicles to go straight and make right and left turns to pass. However, pedestrians are not factored and the allowed crossing time is kept fixed, similar to the first experiment. Finally, the third experiment realizes a fully dynamic TLS, where the needed time for pedestrians to cross has also been considered, as outlined in Figure 16. For that, PCS becomes handy in estimating the required time.

Here we have recreated the traffic scenario using the data we collected in our field study. Figure 17 compares the results. The second and third experiments apply adaptive TLS and thus yield much better performance than static traffic light systems. The vehicle average waiting time in the second and third experiments has dropped, respectively, by 65.04 and 66.02% compared to the first experiment. Similarly, the pedestrian average waiting time has decreased by 75.31 and 77.44%, respectively. The only difference between the second and third experiments is whether PCS is incorporated to factor pedestrians in determining the TLS.

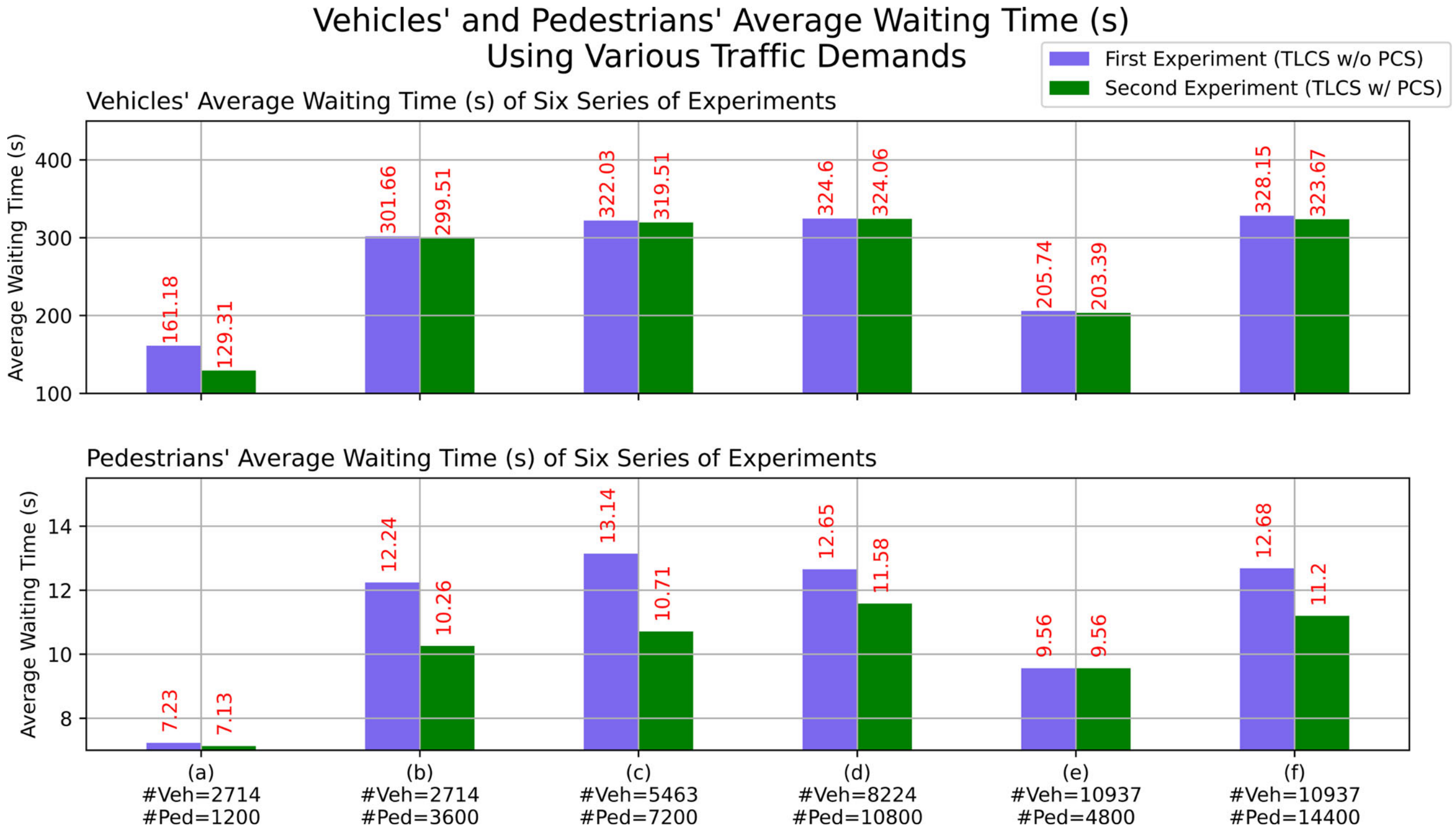

**Figure 19.** Vehicle and pedestrian average waiting time in seconds under various traffic demands.

### *Results using real world traffic demands*

Figure 18 shows the violin plots of the distributions of vehicle and pedestrian waiting time; their max waiting times are annotated as well. In a violin plot, a wider area means higher data density. For example, the leftmost bar in the bottom plot shows the distribution of pedestrian waiting time in our first experiment. The volume of pedestrians who waited for about 48 and 80 s are the largest, as indicated by the width of the bar. The bar above 100 s is very narrow meaning that very few pedestrians waited for longer than 100 s. The TLS with PCS (the third experiment) significantly reduces the max waiting time for both vehicles and pedestrians, where the bar is the shortest among the three experiments. By factoring in pedestrians, enough time is allotted for road crossing while calculating the duration for the next phase. This improves intersection crossing safety; yet, the maximum vehicle waiting time is reduced by not overestimating what constitutes safe crossing time. These improvements illustrate that PCS plays an important role in improving the performance of the TLS.

### *Experiments with various traffic demands*

In our field study, 990 vehicles and 522 pedestrians were involved. In order to show the performance of using PCS for TLS under different traffic demands, six more simulation experiments have been conducted. The vehicle and pedestrian trips have been generated randomly. The starting point and destination is a random pick of the four endpoints in the intersection map, that is, the points on the north/-east/south/west sides of the intersection and which are all 200 m away from the intersection. Figure 19 compares the average waiting time of vehicles and pedestrians. Figure 19(a) and (b), as well as (e) and (f), show that with stable vehicular demand and rising pedestrian density, the vehicle average waiting time grows significantly. This is attributed to the increased delay for vehicles that turn right or left where they have to yield to frequent road-crossing pedestrians. In both scenarios the TLS system with PCS decreases both vehicle and pedestrian average waiting times. Figure 19(b–d, f) show that as vehicular and pedestrian demands grow at the same time, both vehicle and pedestrian average waiting time rises accordingly. The TLS system with PCS, however, significantly decreases pedestrian average waiting time without impacting vehicles; in fact, PCS even reduces the vehicle average waiting time a little bit. All the aforementioned experiment results demonstrate that our proposed simulator contributes to improving the TLS performance in addition to boosting pedestrian safety by scheduling enough time for them to cross, yet without impacting the vehicular traffic performance much.

## Conclusion

Simulating the behavior of pedestrians, and their interaction with vehicles, have attracted lots of attention in recent years. Studying pedestrian motion at intersection crossing would be invaluable for reducing delay and maximizing safety. However, most existing work has focused on road crossing while assuming non-signalized crosswalks. This article has presented PCS, a novel pedestrian road-crossing simulator that captures the behavior of pedestrian groups at signalized intersections. The simulation opts to estimate the needed time to cross and gauge the impact of various parameters for the intersection, for example, crosswalk width, and for the user, for example, age and mobility. Our experiments using both real-world traffic demands and random traffic volume have demonstrated that PCS enables optimized signal scheduling and consequently improves the pedestrian experience in terms of waiting time and enables provisioning sufficient time for safe crossing. PCS is also open source and can be used for other research studies. However, our current PCS implementation is for a single intersection. In the future, we will study the pedestrian behavior by correlating data at multiple adjacent intersections, and devise algorithms to improve the throughput of both vehicles and pedestrians for multiple intersections.

## Disclosure Statement

No potential conflict of interest was reported by the author(s).

## ORCID

Dayuan Tan http://orcid.org/0000-0001-6709-4556
Mohamed Younis http://orcid.org/0000-0003-3865-9217
Wassila Lalouani http://orcid.org/0000-0002-0801-5827